\documentclass[fleqn,twoside]{article} 
\usepackage{espcrc2} 
\input epsf.sty
 
\newcommand{\AmS}{{\protect\the\textfont2 
  A\kern-.1667em\lower.5ex\hbox{M}\kern-.125emS}} 
 
\title{Hadronic structure in $\tau^- \rightarrow
\pi^-\pi^-\pi^+\nu_\tau$ decays} 
 
\author{Edward I. Shibata\address{Department of Physics, Purdue
University,
\\ West Lafayette, Indiana 47907-2036,  United States of America}
\thanks{representing the CLEO Collaboration.}}              
\begin{document} 
 
\begin{abstract} 
	
	A model-dependent analysis of the hadronic substructure in $\tau^- \rightarrow
\pi^-\pi^-\pi^+\nu_\tau$ decay is reported. The decay is
dominated by the process $\tau^- \rightarrow a_1^-(1260) \nu_\tau,$
in which $a_1^-(1260) \rightarrow \rho^0\pi^-$ in an $S$-wave
decay. Amplitudes involving $a_1(1260)$ decays into isoscalars, 
especially $a_1(1260)^- \rightarrow f_0(600) \pi^-,$ are large.  $\tau^- \rightarrow
\pi^-\pi^-\pi^+\nu_\tau$ decays
via the pseudoscalar $\pi^-(1300)$ are small. These results
support the resonant substructure reported in the previously reported
analysis of $\tau^- \rightarrow \pi^0\pi^0\pi^-\nu_\tau$ decay mode from the 
same CLEO~II data sample.
\vspace{1pc} 
\end{abstract} 
 
\maketitle 
 
\section{INTRODUCTION} 
 The  $\tau^- \rightarrow (3\pi)^- \nu_\tau$ decay has been the subject of much interest 
over the years. Because of the transformation properties of the weak current under
parity and $G$-parity, $\tau$-lepton decay to an odd number of pions is expected to occur
exclusively through the axial vector current, ignoring isospin-violating effects. 
Since $J^P,$ and $I^G$ are good quantum numbers of the hadronic current from weak decays,
the weak axial current produces systems of pions with $J^P =0^-$ or 
$1^+,$ odd $G$-parity and odd numbers of pions.

	The $(3\pi)^-$ system in $\tau^- \rightarrow (3\pi)^- \nu_\tau$ is expected to 
be produced dominantly through the poorly understood $J^P = 1^+$  $a_1^-(1260)$ where 
$a_1^-(1260) \rightarrow \rho^0\pi^-$ in an $S$-wave decay. Although it is possible that
the $(3\pi)^-$ system can be produced through the $J^P = 0^-$ $\pi^-(1300),$
this is expected to be suppressed by the PCAC (partially conserved axial current) hypothesis.
It has been known for some time that simple models do not provide a 
satisfactory description. For example, ARGUS \cite{Albr95} found that the model of
K\"{u}hn and Santamaria \cite{Kuhn90} and that of Isgur, Morningstar, and Reader \cite{Isgu89} 
did not describe their $\tau^- \rightarrow (3\pi)^- \nu_\tau$ data satisfactorily. 

	Here we report the preliminary results of a model-dependent fit of the substructure in $\tau^-
\rightarrow
\pi^-\pi^-\pi^+\nu_\tau$ decays. The work is based on the analysis of $\sim 145,000$
$\tau^- \rightarrow \pi^-\pi^-\pi^+\nu_\tau$ events \cite{Hins01}.
In addition to fitting the substructure with various $a_1^-(1260)$ decay modes, fits
for the production of the $(3\pi)^-$ system via the $J^P = 0^-$ 
$ \pi^-(1300)$ are included.
These results are compared to the substructure
in $\tau^- \rightarrow
\pi^0\pi^0\pi^-\nu_\tau$ previously reported \cite{Asne99}.

	Throughout this paper the charge conjugate versions of decays 
are implied for brevity and clarity.

\section{EVENT SELECTION}

\begin{figure}[h]
\centering\leavevmode
  \epsfxsize=7.5cm
  \epsfbox{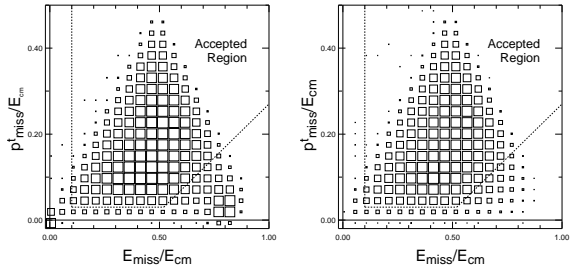}
\caption{$E_{miss}/E_{cm}$ vs. $p^t_{miss}/E_{cm}$ for experimental data (left) and Monte Carlo
$\tau$-events (right), where $E_{miss}$ is the missing energy and $p_{miss}^t$ is
the missing transverse momentum in an event. This
2-dimensional cut is used to reject events with missing tracks and/or energy along the beam axis. }
\label{twodcut}
\end{figure}

	 This analysis is based on 4.67~fb$^{-1}$ of $e^+e^-$ colliding beam data collected at 
center-of-mass energy energies $E_{cm} \sim 10.6$~GeV with the CLEO~II detector at the
Cornell Electron Storage Ring (CESR). This corresponds to $4.3
\times 10^6$ 
$e^+e^- \rightarrow \tau^+\tau^-$ events.
$\tau^- \rightarrow \pi^-\pi^-\pi^+\nu_\tau$
events are selected from a topology in which the 
$\tau^+$ decays into a single charged particle plus a $\bar \nu_\tau$ and the $\tau^-$ decays into
three charged particles plus a $\nu_\tau$ (1 versus 3 charged-particle topology). Selected events are
required to have four well-constructed charged tracks coming from the interaction region, and the net
charge of the tracks is required to be zero.
No more than one track is allowed to be identified as an $e^-$ or $e^+.$
The most isolated track is identified, and the remaining three tracks are required
to be $\ge 90^\circ$
away from it.
Events with photons (isolated electromagnetic showers not associated with a track) are rejected.
Events that satisfy the hypothesis $0.485 < M_{\pi^+\pi^-} < 0.510$~GeV are rejected in order
to eliminate events containing $K_s^0 \rightarrow \pi^+\pi^-$ decays. 
Finally, the 2-dimensional cut shown in Figure~\ref{twodcut} is used to reject events with missing tracks
and/or energy along the beam pipe. Application of these cuts  yields
145,000~$\tau^- \rightarrow \pi^-\pi^-\pi^+\nu_\tau$ events.

\begin{figure}[ht]
\centering\leavevmode
  \epsfxsize=7cm
  \epsfbox{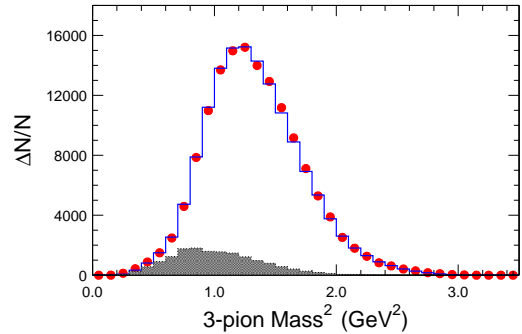}
\caption{$\pi^+\pi^+\pi^-$ effective mass spectrum. The circles
are the experimental data and
the histogram is the spectrum resulting from fitting the 
data sample as described in the text. The shaded histogram is the background from other
$\tau$ decays.}
\label{threepimass}
\end{figure}

	The $\pi^-\pi^-\pi^+$ effective mass spectrum resulting from these cuts is shown in
Figure~\ref{threepimass}.  The background, indicated by the shaded histogram,
which  amounts to about 11\% of the events
in the spectrum. Monte Carlo studies show that this background is from other
$\tau$-decays: $\sim {1 \over 2}$ comes from $\tau$-events with missing $\pi^0$('s) and 
$\sim {1 \over 2}$ is from $\tau$-events with $\ge 1$ charged $K.$  Monte Carlo studies
indicate that backgrounds from  multihadronic $q \bar q$ events, $\gamma\gamma$ events,
and QED events are
negligible.

\section{FIT}

\begin{table*}[t]
\caption{Masses and widths of the resonances used in the fits to $\tau^- \rightarrow
\pi^-\pi^-\pi^+\nu_\tau$ and  $\tau^- \rightarrow \pi^0\pi^0\pi^-\nu_\tau$ decays. 
For comparison the current Particle Data Group \cite{Pdg02} masses and widths are listed. }
\label{tableone}
\begin{tabular}{cccccc}\hline
Particle & $I^G(J^{PC})$ & M (GeV) & $\Gamma$ (GeV) & $M_{PDG}$ (GeV) & $\Gamma_{PDG}$ (GeV)\\ \hline
$a_1(1260) $ & $1^-(1^{++})$ & 1.230 & 0.400 & $1.23 \pm 0.04$ &  0.25--0.60  \\
$\pi(1300) = \pi^\prime$ & $1^-(0^{-+})$ & 1.300 & 0.300 & $1.3 \pm 0.1$ &  0.2--0.6  \\ 
$\rho(770)$ & $1^+(1^{--})$ & 0.774 & 0.149 & $ 0.7711 \pm 0.0009$ & $0.1492 \pm 0.0007$ \\
$\rho(1450)$ & $1^+(1^{--})$ & 1.370 & 0.386 & $1.465 \pm 0.025$ & $0.31 \pm 0.06$\\
$f_2(1270)$ & $0^+(2^{++})$ & 1.275 &  0.185 & $1.2754 \pm 0.0012$ & $0.1851^{+0.0034}_{-0.0026}$ \\
$f_0(600) = \sigma$ & $0^+(0^{++})$ & 0.860 & 0.880 &   0.4--1.2  & 
 0.6--1.0  \\
$f_0(1370)$ & $0^+(0^{++})$ & 1.186 & 0.350 &  1.2--1.5  &  0.2--0.5  \\
\hline
\end{tabular}
\end{table*}

	Although it is desirable to characterize the structure of the $\pi^-\pi^-\pi^+$ system
without model-dependent assumptions, this is very  difficult in practice. In a model-independent
approach one would assume a general form of the hadronic current that allows the separation
and analysis of the complex magnitudes of the axial vector and pseudoscalar components
in bins of $Q^2,$ $s_1,$ and $s_2,$ where $Q^2$ is the invariant mass-squared of the 
$\pi^-\pi^-\pi^+$
system, $s_1$ is the invariant mass-squared of the $\pi^+\pi^-$ combination that yields the larger value
of mass-squared,
and $s_2$ is the invariant mass-squared of the other $\pi^+\pi^-$ combination. There would be
no assumptions regarding resonances in the decay. However, in certain chiral limits,
scalar effects can be induced in the current from the $a_1^-(1260)$ and non-resonant
contributions in the current and thus fake pseudoscalar effects \cite{Deck94}. Secondly,
simply binning in three dimensions is problematical: relatively small bins are
required to study the $\tau$ decay structure and this leads to some bins with
low statistics that may be crucial in resolving complicated resonance structure.
By using a model-dependent fit, one can make reasonable assumptions about
the resonant structure and study the likely existence of  each resonance modeled while using the entire
dataset in the fitting and without the need for binning.

	Using an unbinned maximum likelihood fit, the substructure in $\tau^- \rightarrow
\pi^-\pi^-\pi^+\nu_\tau$ decays has been fit with the set of resonances given in Table~\ref{tableone}.
For comparison purposes, the current Particle Data Group \cite{Pdg02} masses and widths are
listed. For the 
$f_0(600)$ and $f_0(1370)$ resonance parameters given by 
the T\"{o}rnqvist unitarized quark model \cite{Torn95} have been used. 
The same resonances and parameters have been used in the analysis of $\tau^- \rightarrow
\pi^0\pi^0\pi^-\nu_\tau$ events
\cite{Asne99} selected from the same CLEO~II set of $e^+e^- \rightarrow \tau^+\tau^-$ events.

\section{RESULTS}

The results of the 
fit are given in Table~\ref{table:2}, where 
the $\beta_i $ are complex coupling constants. 
The $\pi^-\pi^-\pi^+$ invariant mass spectrum resulting from
the fit is shown as the histogram in Figure~\ref{threepimass}. 

\begin{figure*}[pt]
\centering\leavevmode
  \epsfxsize=14cm
  \epsfbox{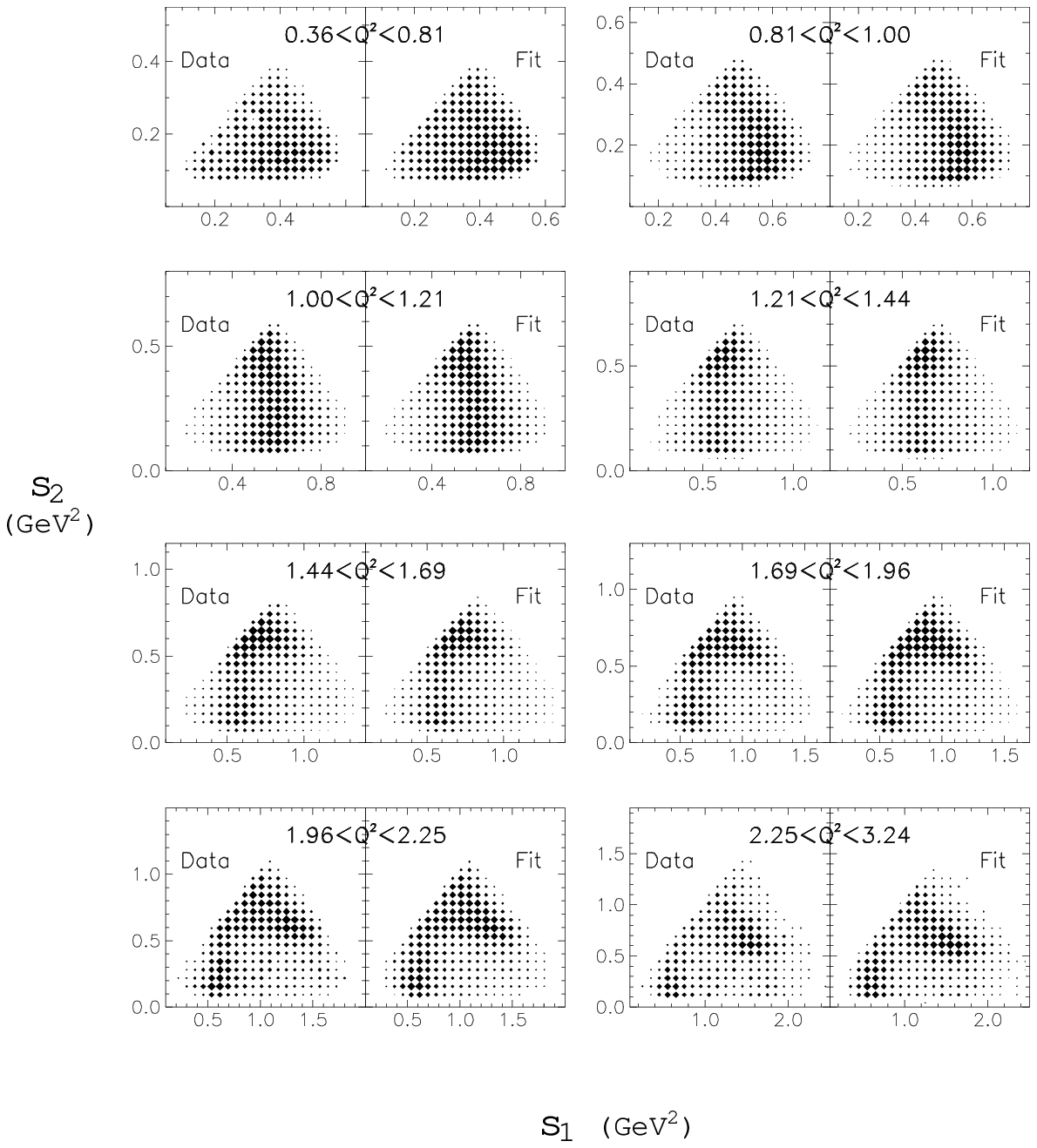}
\caption{Dalitz plots for sequential bins in $Q^2 = M_{\pi^-\pi^-\pi^+}^2.$ $Q^2$ is in units 
of (GeV)$^2.$ $s_1$ and
$s_2$ are the two possible $\pi^+\pi^-$ invariant masses squared; $s_1$ is 
the larger of the two possible $\pi^+\pi^-$ invariant mass-squared values. }
\label{dalitz}
\end{figure*}

	As expected, the $a_1^-(1260) \rightarrow \rho^0\pi^-$
$S$-wave  decay dominates. There is a significant amount of $p$-wave $a_1(1260)$
decay into isoscalars [$f_0(600),$ $f_0(1370)$] plus a pion. $\tau$-decay
via the pseudoscalar $\pi^-(1300)$ is small, although its inclusion improves
the fit. Dalitz plots for sequential  $\pi^-\pi^-\pi^+$ invariant mass regions
are shown in Figure~\ref{dalitz} for data and the fit.

\begin{table*}[h]
\caption {Preliminary results based on $\sim 145,000$ $\tau^- \rightarrow \pi^-\pi^-\pi^+\nu_\tau$
events. The uncertainties are statistical only. The branching fractions do not sum to 100\% due to
interferences among the amplitudes. The phases are relative to the dominant $a_1(1260) \rightarrow
\rho\pi$
$s$-wave  decay.}
\label{tabletwo}
\begin{tabular}{llccc}\hline
	& & Branching fraction (\%) & ${\cal R}$e($\beta$) & ${\cal I}$m($\beta$) \\ \hline
$a_1 \rightarrow \rho\pi$ & $S$-wave &69.77 &1.00 &0.0\\
$a_1 \rightarrow \rho(1450)\pi$ & $S$-wave & $1.58 \pm 0.25$ & $0.05 \pm 0.02$ &$-0.25 \pm 0.02$\\
$a_1 \rightarrow \rho\pi$ & $D$-wave & $1.68 \pm 0.16$ & $0.61 \pm 0.03$ & $0.35 \pm 0.05$\\
$a_1 \rightarrow \rho(1450)\pi$ & $D$-wave & $2.77 \pm 0.29$ &$-1.60 \pm 0.11$ &$-1.21 \pm 0.10$\\
$a_1 \rightarrow f_2(1270)\pi$ & $P$-wave & $0.68 \pm 0.08$ &$-0.12 \pm 0.07$ &$0.96 \pm 0.06$\\
$a_1 \rightarrow f_0(600) \pi$ & $P$-wave &  $48.40 \pm 1.61$  &  $1.99 \pm 0.06$ 
 &$2.70 \pm 0.06$ \\
$a_1 \rightarrow f_0(1370)\pi$ & $P$-wave & $11.51 \pm 0.82$ &$-0.04 \pm 0.04$ &$-1.15 \pm 0.04$\\ 
$\pi(1300) \rightarrow \rho\pi$ & $P$-wave & $0.01 \pm 0.02$  & $0.001 \pm 0.002$ & $-0.002 \pm 0.001$\\
$\pi(1300) \rightarrow \rho(1450)\pi$ & $P$-wave & $0.02 \pm 0.03$  & $-0.005 \pm 0.010$ & $-0.013 \pm
0.009$\\
$\pi(1300) \rightarrow f_0(600) \pi$ & $S$-wave & $0.13 \pm 0.07$  & $-0.005 \pm 0.003$ & $-0.008 \pm
0.002$\\ \hline
\end{tabular}
\end{table*}

\begin{table*} [b]
\caption{Results for the substructure fit to $ \sim 15,000$ lepton-tagged $\tau^- \rightarrow
\pi^0\pi^0\pi^-\nu_\tau$ events from the same CLEO~II $e^+e^- \rightarrow \tau^+ \tau^-$ dataset.
The branching fractions do not sum to 100\% due to
interferences among the amplitudes.} 
\label{table:2}
\tabcolsep 4pt 
\begin{tabular}{llcccc} 
\hline
& &  Significance & Branching fraction~(\%) &  $\vert\beta\vert$  
&  phase $\varphi /\pi$  
\\ \hline 
%
$\rho\pi$ & $S$-wave & --  & $ 68.11 $ & $ 1.00 $ & $ 0.0  $ \\ 
$ \rho(1450)\pi $ & $S$-wave  
& $1.4\sigma$  & $ 0.30 \pm 0.64 $ & $ 0.12 \pm 0.09 $ &  
                                              $\phantom{-}0.99 \pm 0.25 $ \\ 
$ \rho\pi $       & $D$-wave  
& $5.0\sigma$  & $ 0.36 \pm 0.17 $ & $ 0.37 \pm 0.09 $ &  
                                                        $-0.15 \pm 0.10 $ \\ 
$ \rho(1450)\pi $ & $D$-wave  
& $3.1\sigma$  & $ 0.43 \pm 0.28 $ & $ 0.87 \pm 0.29 $ &  
                                              $\phantom{-}0.53 \pm 0.16 $ \\ 
$ f_2(1270)\pi  $ & $P$-wave  
& $4.2\sigma$  & $ 0.14 \pm 0.06 $ & $ 0.71 \pm 0.16 $ &  
                                              $\phantom{-}0.56 \pm 0.10 $ \\ 
 $f_0(600) \pi $  &  $P$-wave 
&  $8.2\sigma$   
&  $16.18 \pm 3.85 $  &  $ 2.10 \pm 0.27 $  &  
                                             $\phantom{-}0.23 \pm 0.03 $  \\ 

$ f_0(1370)\pi  $ & $P$-wave  
& $5.4\sigma$  & $ 4.29 \pm 2.29 $ & $ 0.77 \pm 0.14 $ &  
                                                        $-0.54 \pm 0.06 $ \\ 
%
%
\hline 
\hline 
\end{tabular} 
\end{table*}

\begin{figure}[h]
\centering\leavevmode
  \epsfxsize=7cm
  \epsfbox{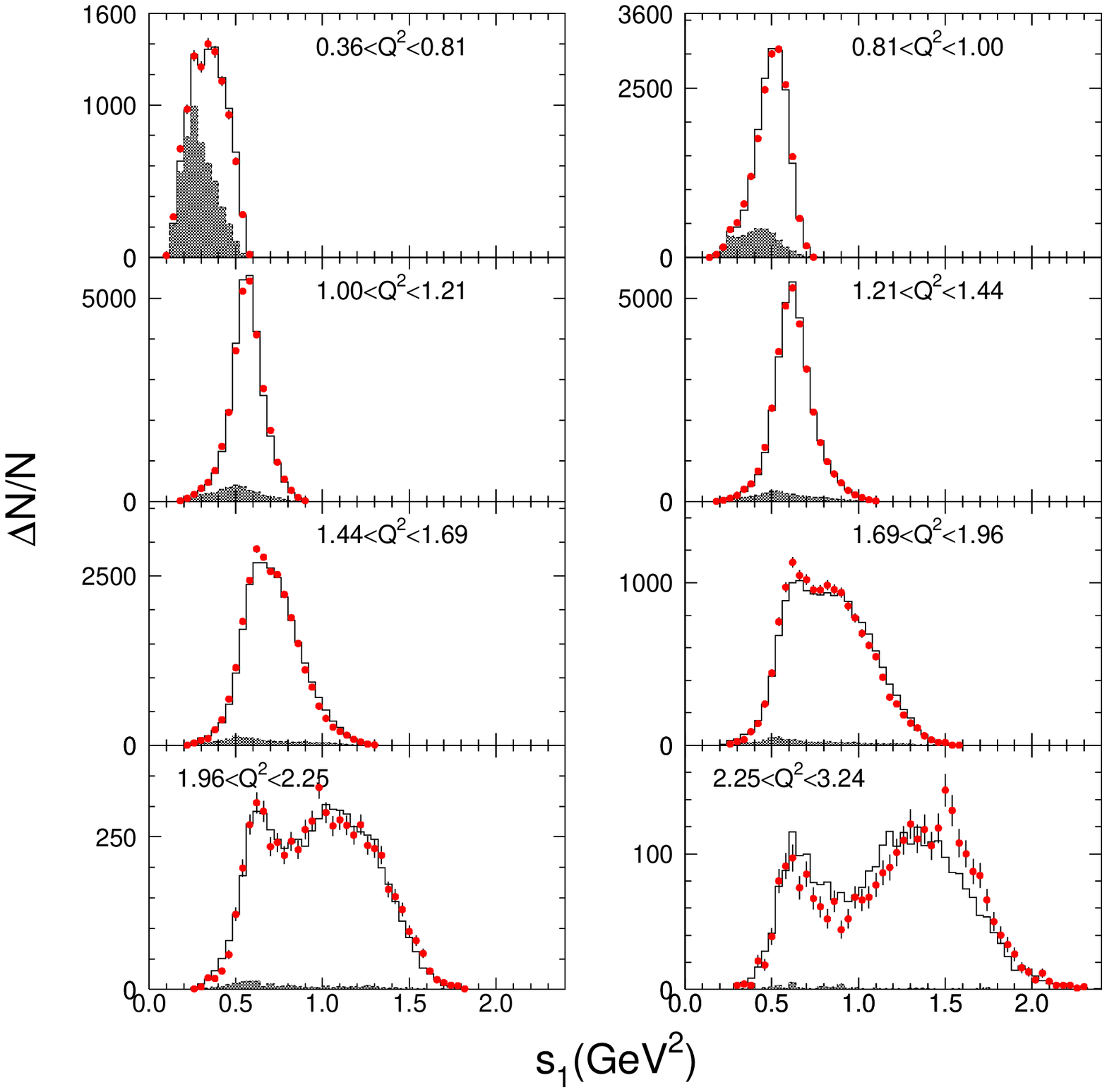}
\caption{Projection of the Dalitz plot onto the $s_1 = M^2_{\pi^+\pi^-}$ axis for various bins in
$M^2_{3\pi}$ for a fit without
isoscalars. The points are the experimental data and the histogram
is the result of the fit. The shaded regions represent the background.}
\label{fig:6}
\end{figure}

\begin{figure}[h]
\centering\leavevmode
  \epsfxsize=7cm
  \epsfbox{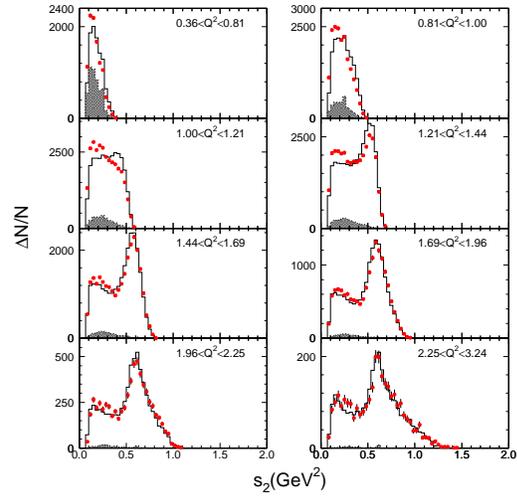}
\caption{Projection of the Dalitz plot onto the $s_2 = M^2_{\pi^+\pi^-}$ axis for various bins in
$M^2_{3\pi}$ for a fit without
isoscalars. The points are the experimental data and the histogram
is the result of the fit. The shaded regions represent the background.}
\label{fig:7}
\end{figure}

The isoscalar sub-resonances in the decay of the $a_1^-(1260)$ [$f_0(600),$ $f_2(1270),$
and $f_0(1370)$] play a large role in the fits. To demonstrate the need for these isoscalars,
the data were fitted without them. Projections of the Dalitz plat without the presence of these
isoscalars are shown in Figures~\ref{fig:6} and \ref{fig:7}.
For the projection onto the $s_1$ axis
(Figure~\ref{fig:6}), the lack of all three isoscalars 
results in a  poor fit for $Q^2 \ge 2.25$~GeV$^2.$
For the projection onto the $s_2$ axis (Figure~\ref{fig:7}), the presence of $f_0(600)$ is needed
for all regions of
$Q^2,$ especially those of lower $Q^2$ values. 

These results support the resonance substructure previously reported for the
CLEO~II $\tau^- \rightarrow \pi^0\pi^0\pi^-\nu_\tau$ \cite{Asne99} data, which are 
based on $\sim 15,000$
lepton-tagged events. The fit results on these events are
given in Table~\ref{table:2} for the reader's convenience. Both analyses find that
axial vector amplitudes with isoscalars, especially $a_1^-(1260) \rightarrow f_0(600) \pi^-,$
are large.

\clearpage
\clearpage

\begin{figure}[h]
\centering\leavevmode
  \epsfxsize=7cm
  \epsfbox{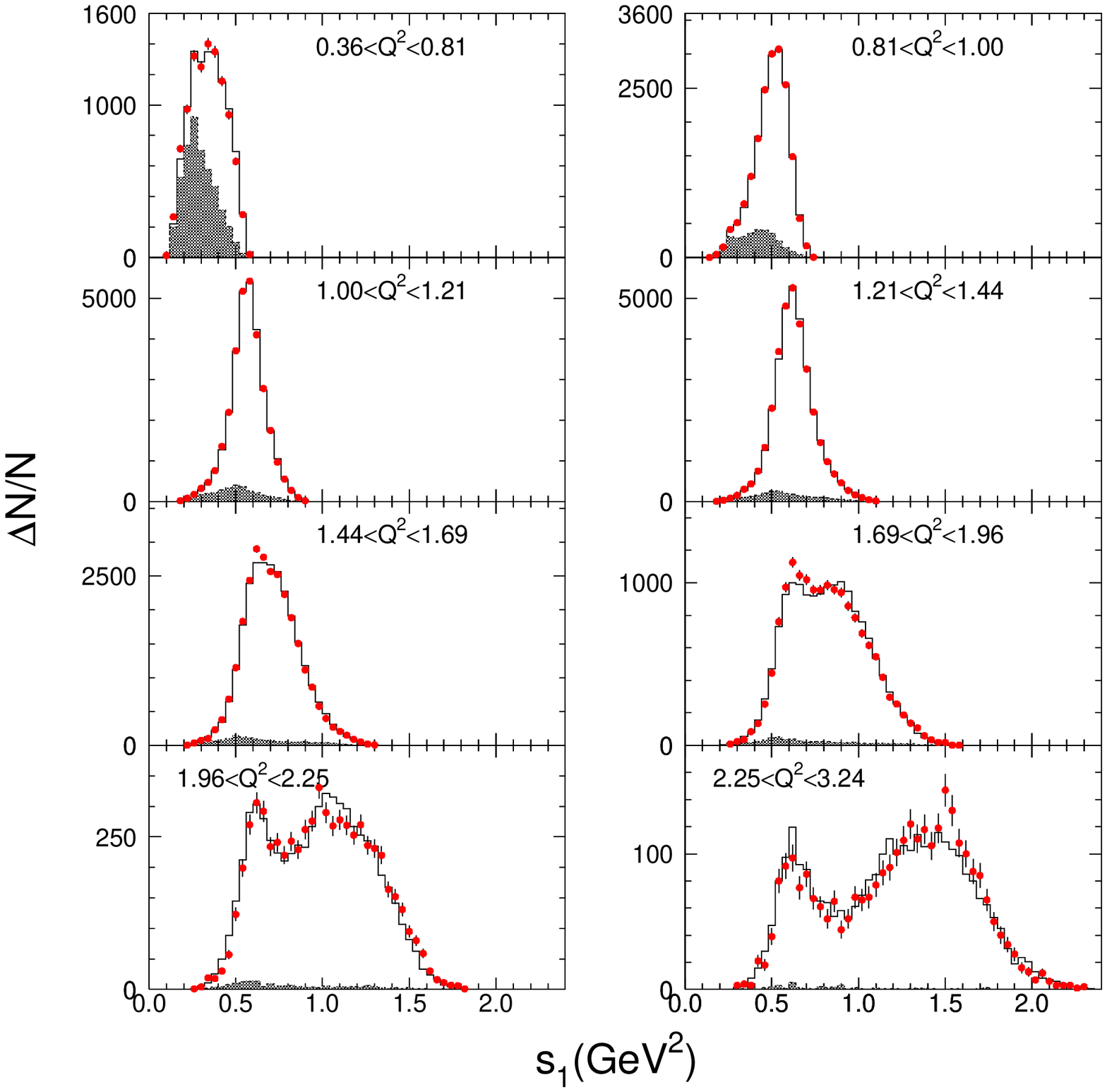}
\caption{Projection of the Dalitz plot onto the $s_1 = M^2_{\pi^+\pi^-}$ axis for various bins in
$M^2_{3\pi}.$ The points are the experimental data and the histogram
is the result of the fit. The shaded regions represent the background.}
\label{soneproj}
\end{figure}

\begin{figure}[h]
\centering\leavevmode
  \epsfxsize=7cm
  \epsfbox{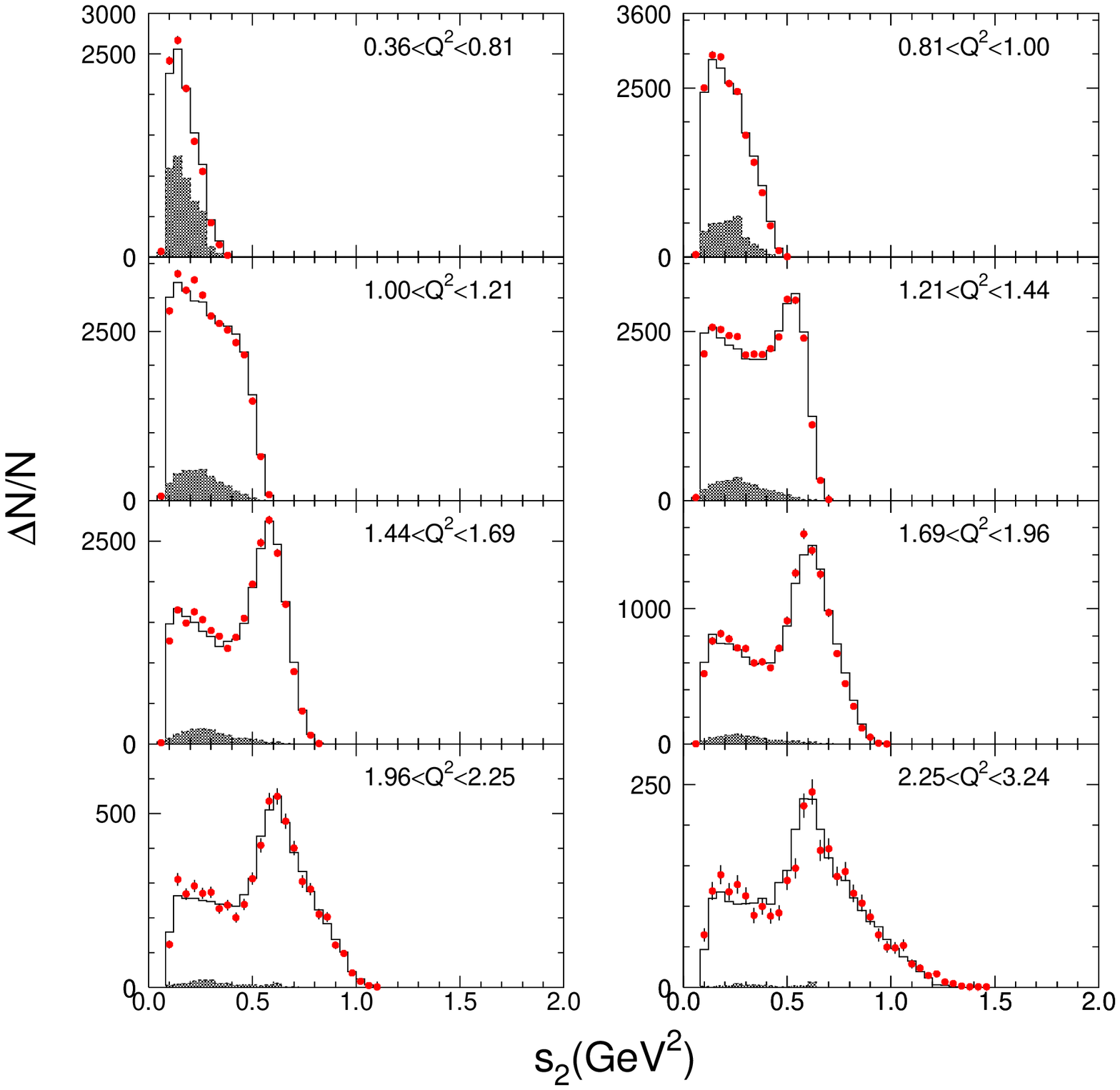}
\caption{Projection of the Dalitz plot onto the $s_2 = M^2_{\pi^+\pi^-}$ axis for various bins in
$M^2_{3\pi}.$ The points are the experimental data and the histogram
is the result of the fit. The shaded regions represent the background.}
\label{fig:5}
\end{figure}

\section{SUMMARY}
  This work is a new measurement of substructure in $\tau^-\rightarrow
\pi^-\pi^-\pi^+\nu_\tau. $ A model-dependent unbinned maximum likelihood
fit on 145,000  events was performed using an axial vector component [$a_1(1260)$ primary resonance and
several subresonances, including isoscalars] and a pseudoscalar component [$\pi(1300)$
plus several subresonances].  As expected, this 
decay is dominated by 
$\tau^- \rightarrow a_1^-(1260) \nu_\tau$ with subsequent 
$a_1 \rightarrow \rho^0\pi^-$ $S$-wave decay. However, it is also apparent that other $a_1^-(1260)$ decay
modes are present. In particular, $a_1^-(1260)$ decay into isoscalars [$f_0(600),$ $f_2(1270),$ and
$f_0(1370)$] is significant. The $a_1^-(1260) \rightarrow f_0(600) \pi^-$ is especially prominent. These
results are consistent with the earlier CLEO~II results for
$\tau^-\rightarrow
\pi^0\pi^0\pi^-\nu_\tau$ \cite{Asne99}. Both analyses find that axial-vector amplitudes with isoscalars, 
especially $a_1^-(1260) \rightarrow f_0(600) \pi^-,$ are large.
Regarding non-axial-vector contributions, inclusion of the $J^P = 0^-$
$\pi^-(1300)$ improves the fit to the data in the expected regions, but the need for it is not compelling.
It is hoped that the future CLEO-c dataset will clarify some of these issues.

\section{ACKNOWLEDGEMENTS}
This work represents the work of many CLEO collaborators, especially Jason W. Hinson, Jon Urheim, 
and Alan J. Weinstein. The CESR staff provided us with
excellent luminosity and running conditions. This work is supported by the National
Science Foundation, the U. S. Department of Energy, and the Natural Sciences and
Engineering Research Council of Canada.

\end{document}